\documentclass[twocolumn,preprintnumbers,aps,prl, nofootinbib]{revtex4-2}
\usepackage{amsmath,amssymb,amsthm}
\usepackage{graphicx}
\usepackage{tabularx}
\usepackage{paralist}
\usepackage{epstopdf}
\usepackage{bbold}
\usepackage{braket}
\usepackage{pgfgantt}
\usepackage{lmodern}

\usepackage{color}
\usepackage{bm}

\usepackage{lpic}
\usepackage[symbol]{footmisc}
\usepackage{tikz, tikzscale, pgfplots}
\usetikzlibrary{matrix, fit}
\usetikzlibrary{decorations.pathreplacing,angles,quotes}
\usetikzlibrary{arrows,fit, quotes,arrows.meta, spy, calc, decorations.markings, shapes.misc, positioning, patterns, shadings, decorations.shapes}

\usepackage[utf8]{inputenc}
\usepackage[english]{babel}
\usepackage{amsmath,amssymb,amsthm}
\usepackage{graphicx}
\usepackage{tabularx}
\usepackage{paralist}
\usepackage{epstopdf}
\usepackage{bbold}
\usepackage{braket}
\usepackage{pgfgantt}
\usepackage{centernot}

\definecolor{mygreen}{rgb}{0,.6,0}
\newcommand{\CHANGE}[1]{{{#1}}}
\newcommand{\CHANGEB}[1]{{{#1}}}

\newcommand{\xa}[1]{\begin{tikzpicture}\node[draw,dotted] {$#1$};\end{tikzpicture}}
\newcommand{\xb}[1]{\begin{tikzpicture}\node[draw,dashed] {$#1$};\end{tikzpicture}}
\newcommand{\xc}[1]{\begin{tikzpicture}\node[draw] {$#1$};\end{tikzpicture}}

\renewcommand{\t}[1]{\mathrm{#1}}

\begin{document}

		\title{Parallel selective nuclear spin addressing for fast high-fidelity quantum gates}			
		
%			\author{Benedikt Tratzmiller$^{1, \ast}$, Jan F. Haase$^{2,3}$, Zhenyu Wang$^{1,4,5}$ and Martin B. Plenio$^{1, \ast}$
							\author{Benedikt Tratzmiller$^{1}$, Jan F. Haase$^{2,3}$, Zhenyu Wang$^{1,4,5}$ and Martin B. Plenio$^{1}$
				\\
				\small{$^{1}$ Institut f\"{u}r Theoretische Physik und IQST, Albert-Einstein-Allee 11, Universit\"{a}t Ulm, 89081 Ulm, Germany}\\
				\small{$^{2}$ Institute for Quantum Computing, University of Waterloo, 200 University Avenue West, Waterloo, Ontario, Canada N2L 3G1}\\
				\small{$^{3}$ Department of Physics \& Astronomy, University of Waterloo, Waterloo, ON, Canada, N2L 3G1}\\
				\small{$^{4}$ Guangdong Provincial Key Laboratory of Quantum Engineering and Quantum Materials,  School of Physics and Telecommunication Engineering, South China Normal University, Guangzhou 510006, China}
				\\
				\small{$^{5}$ 
					\CHANGE{Guangdong-Hong Kong Joint Laboratory of Quantum Matter, }
					Frontier Research Institute for Physics, South China Normal University, Guangzhou 510006, China}
				\\
			}

%		}

		\date{\today}
		
\begin{abstract}
	Due to their long coherence times, nuclear spins have gained considerable attention as physical qubits.
Two-qubit gates between nuclear spins of distinct resonance frequencies can be mediated by electron spins,
usually employing a sequence of electron-nuclear gates. Here we present a different approach inspired by, but not limited
to, NV centers in diamond and discuss possible applications. To this end we generalize external electron
spin control sequences for nuclear spin initialization and hyperpolarization to achieve the simultaneous control
of distinct nuclear spins via an electron spin. This approach results in efficient entangling gates that,
compared to standard techniques, reduce the gate time by more than 50\% when the gate time is limited by off-resonant 
coupling to other spins, and by up to 22\% when the gate time is limited by small electron-nuclear coupling.	

		\end{abstract}
		
		\maketitle

\section{I. Introduction}

%\footnote[0]{$^\ast$ benedikt.tratzmiller@uni-ulm.de; martin.plenio@uni-ulm.de}
Spin carrying material impurities represent a promising platform for various near term applications of quantum technology. In 
particular, the nitrogen-vacancy (NV) center has proven itself to be a valid candidate for quantum computation, communication 
and sensing applications \cite{doherty2013nitrogen, staudacher2013nuclear, kalb2017entanglement, wu2016diamond}. Recent progress manifests in a number of works, including NV-nuclear 
quantum gates \cite{jelezko2004observation}, NV-NV gates \cite{dolde2013room}, sensing of single nuclei \cite{muller2014nuclear}
and polarization of spin ensembles \cite{london2013detecting,PP}. However, the coherence time of NV-centers 
limits achievable fidelities. Nuclear spins in diamond may be fully controlled by NV centers and offer far longer coherence 
times due to their smaller gyromagnetic ratio which renders them less susceptible to environmental noise and thus as ideal candidates 
for physical qubits, e.g. for quantum memories with long storage times. 

Numerous approaches to realize gates between NV centers and nuclear spins have been presented, both with \cite{bradley201910, wang2017delayed} 
and without additional external radiofrequency (RF) control \cite{taminiau2014universal, ZyJg,casanova2016noise} on the nuclear 
spins themselves. Their common foundation are pulsed dynamical decoupling sequences which exhibit resonance frequencies that are multiples 
of $\pi/\tau$ where $\tau$ is the time between subsequent pulses. At present, these sequences are tuned to a single resonance which
implies the major drawback that only one nuclear spin can be controlled at a time.

In this work, we overcome this 
limitation with an approach based on polarization sequences \cite{PP} that originally have been developed to initialize nuclear spins \cite{london2013detecting, scheuer2016optically}. Our modifications allow for simultaneously resonant addressing of two different frequencies at the same time without requiring additional RF control, using a sequence with two free parameters incorporated by a pulse spacing and pulse phases.
This paves the way for a wide range of applications, such as the realization of gates for quantum computation applications or the protection from magnetic field fluctuations.
These are not limited to diamond based material, for example carbon-13 and silicon-29 spins with silicon vacancies in silicon carbide \cite{radulaski2017scalable} can be controlled jointly.
The approach presented here can be used to initialize and manipulate different nuclear spin species simultaneously, extending approaches for quantum simulators made of NV-controlled nuclei \cite{Cai2013}.

We start by explaining the basic pulse sequences that create the desired Hamiltonian and continue with modifications that allow to manipulate the effective coupling strength in this Hamiltonian. These ideas are then applied to investigate the efficiency with which our protocol can create nuclear spin entanglement in two relevant systems, namely between a silicon-29 and a carbon-13 spin like inside silicon-carbide and between carbon-13 spins close to an NV-center in diamond that only differ by their hyperfine coupling to the electron spin. Furthermore we discuss the manipulation of states that are insensitive to magnetic field fluctuations, which therefore offer excellent coherence properties for quantum sensing and computation applications. Finally we compare our approach to standard techniques for different experimentally relevant situations.

%\newpage

\section{II. Tuning Pulsed Polarization sequences resonant to two arbitrary frequencies}

We assume an NV center placed at $\vec{r}=\vec{0}$ in a 
magnetic field $B(\vec{r})=B_z(\vec{r})\,\hat{z}$, which is aligned with its symmetry axis $\hat{z}$.
Under the application of an external microwave drive $H_\t{mw}(t)$, the total Hamiltonian then reads
\begin{align} \label{eq:NV_Ham_complete}
	H(t) =& DS_z^2 {+} \gamma_e B_z(\vec{0}) S_z {+} \sum_n \gamma_n B_z(\vec{r}_n) I_z^n \nonumber \\
	&+ S_z \sum_n \vec{A}_n \vec{I}_n + H_\t{mw}(t),
\end{align}
where $S_z$ is the spin-1 $z$-operator of the NV and we have the nuclear spin-$1/2$ operators $I_i^{n} = \sigma_i^{n}/2$, $i=x,y,z$ \CHANGE{of the n-th nuclear spin} with $\sigma_i$ the corresponding Pauli matrix.
Furthermore, we have the electronic (nuclear) gyromagnetic ratio $\gamma_e$ ($\gamma_n$).
The large zero field splitting $D=2\pi \times 2.87\,\t{GHz}$ permits application of the secular approximation, hence the coupling of each nuclear spin to the electron spin is solely determined by the hyperfine vector $\vec{A}_n$ given by
\begin{align}
	\vec{A}_n = \frac{\mu_0 \gamma_e\gamma_n}{4\pi |\vec{r}_n|^3} \left[\hat{z} - 3 \frac{(\hat{z}\cdot \vec{r}_n)\vec{r}_n}{|\vec{r}_n|^2}\right],
\end{align}
where $\vec{r}_n$ describes the position of the corresponding nucleus.
We now reduce the electronic Hilbertspace to the space spanned by $|m_s=0\rangle$ and $|m_s=+1\rangle$, effectively choosing our working qubit.
Adapting the microwave control to drive transitions between those levels, it takes the form
\begin{align}
	H_{\t{mw}}(t) = 2\Omega(t) \cos(\omega_\t{mw} + \varphi) \sigma_x,
\end{align}
where $\Omega(t)$ is a time dependent Rabi frequency and $\sigma_x =|0\rangle\langle +1|+|+1\rangle\langle 0|$.
In a frame rotating with respect to the zero field splitting and the microwave frequency $\omega_\t{mw}$, the total Hamiltonian can be written as $H=H_0+H_\t{mw}$, where
\begin{align}\label{eq4}
	H_0 =&  \sum_n \vec{\omega}_n {\cdot} \vec{I}^{n} + \frac{1}{2}\sigma_z \sum_n \vec{A}_n  {\cdot} \vec{I}_n,  \\
	H_\t{mw} =&\Delta \sigma_z  /2 + \Omega(t) \left[\sigma_x \cos(\varphi) + \sigma_y \sin(\varphi)\right]  /2 \nonumber
\end{align}
and the strongly oscillating terms are neglected.
Here we defined $\sigma_z =|+1\rangle\langle +1|-|0\rangle\langle 0|$ and $\sigma_y =i|0\rangle\langle +1|-i|+1\rangle\langle 0|$.
The detuning from the electronic transition frequency is denoted as $\Delta$ and the phase $\varphi$ of the drive selects the axis around which the qubit's Bloch vector rotates.
Note that the Larmor vector of the nuclear spins is now further modified by the hyperfine coupling as
\begin{align}
	\vec{\omega}_n = \gamma_n B_z(\vec{r}_n) \hat{z} + \frac{1}{2}\vec{A}_n.
\end{align}
This shift originates from the reduction \CHANGE{of the three-level NV spin-1} to the two-level system and
can be understood as 
an additional magnetic field created by the electron spin, with its magnitude being the average of the vanishing field in $\ket{0}$ and the field $\vec{A}_n$ in $\ket{1}$\CHANGE{, see also \cite{taminiau2012detection}}.
In particular, this is the reason why single nuclear spin addressing becomes possible, even for identical species and homogeneous magnetic fields.

Changing the basis such that the Larmor term $\vec{\omega}_n$ defines the new z-axis and the new y-axis is chosen such that the Hamiltonian is
\begin{equation}\label{eq6}
	H_0 = \sum_n \omega_n I_z^{n} + \frac{1}{2}\sigma_z \sum_n \left(a_\perp I_x^n + a_\parallel I_z^n\right),
\end{equation}
where $\vec{\omega}_n = \omega_n \hat{\omega}_n$, $a_\parallel = \hat{\omega}_n \vec{A}_n$ describes the coupling parallel to the Larmor vector and $a_\perp = \sqrt{|\vec{A}_n|^2 - a_\parallel^2}$ is the orthogonal component.

Instead of the \CHANGE{naturally} occurring interaction in \CHANGE{Eq.~}(\ref{eq6}) that, without additional microwave control $\Omega(t)=0$, conserves the energy of the electronic spin, we now aim to create a polarization exchange interaction of the form $\sigma_x I_x^n\pm \sigma_yI_y^n$.
This can be achieved by applying a sequence of control pulses on the electronic spin that we will detail in the following.

Sensing sequences consisting of equally spaced population inverting pulses like the Carr-Purcell-Meiboom-Gill (CPMG) sequence \cite{carr1954effects} or the XY-family \cite{maudsley1986modified} create an effective $\sigma_z I_{x,y}$ interaction between the electron spin qubit and nuclei with Larmor frequency $\omega_j$ by choosing the 
delay between pulses $\tau$ according to $\tau = k \pi/\omega_j$, where $k$ is an odd integer.
Note that delaying the start of the sequence with respect to $t=0$ effectively chooses the nuclear spin interaction operator $I_{x,y}$, i.e. a convex combination of $I_x$ and $I_y$.
Further, additional pulses can map the effective electronic operator, e.g. a $\pi/2$ pulse transforms the interaction into $\sigma_x I_{x,y}$.

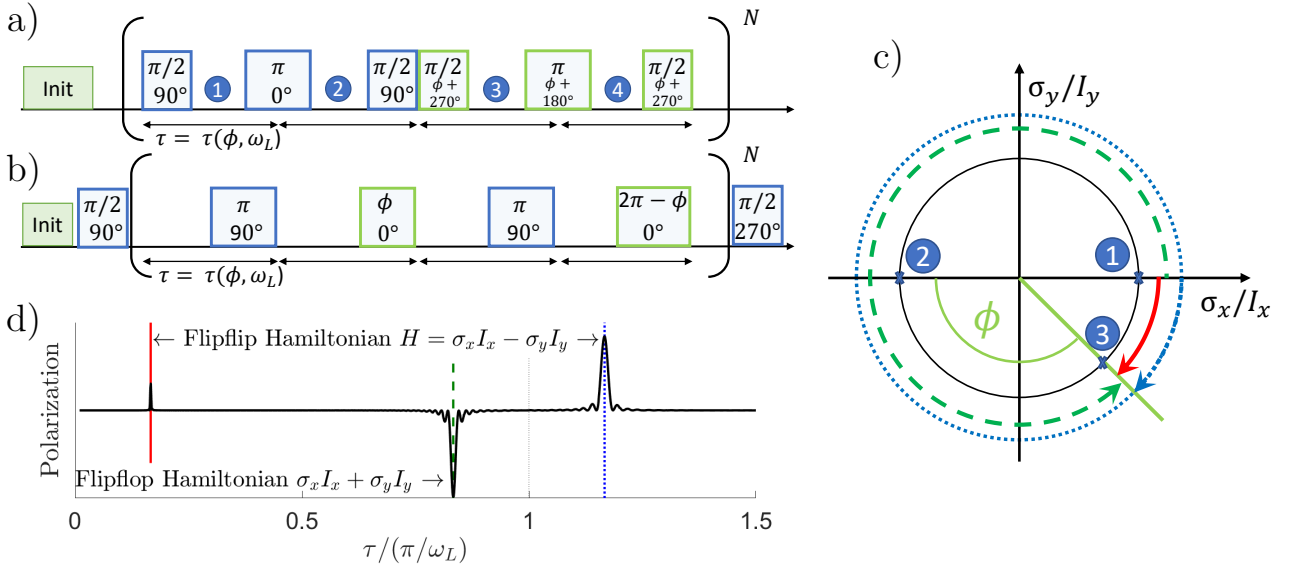
\begin{figure*}[ht]
	\centering
	\begin{tikzpicture}
	\node[inner sep=0pt] (Aa) at (0,0)
	{\includegraphics[width=.6\textwidth, trim=0cm 3cm 11.5cm 4cm, clip]{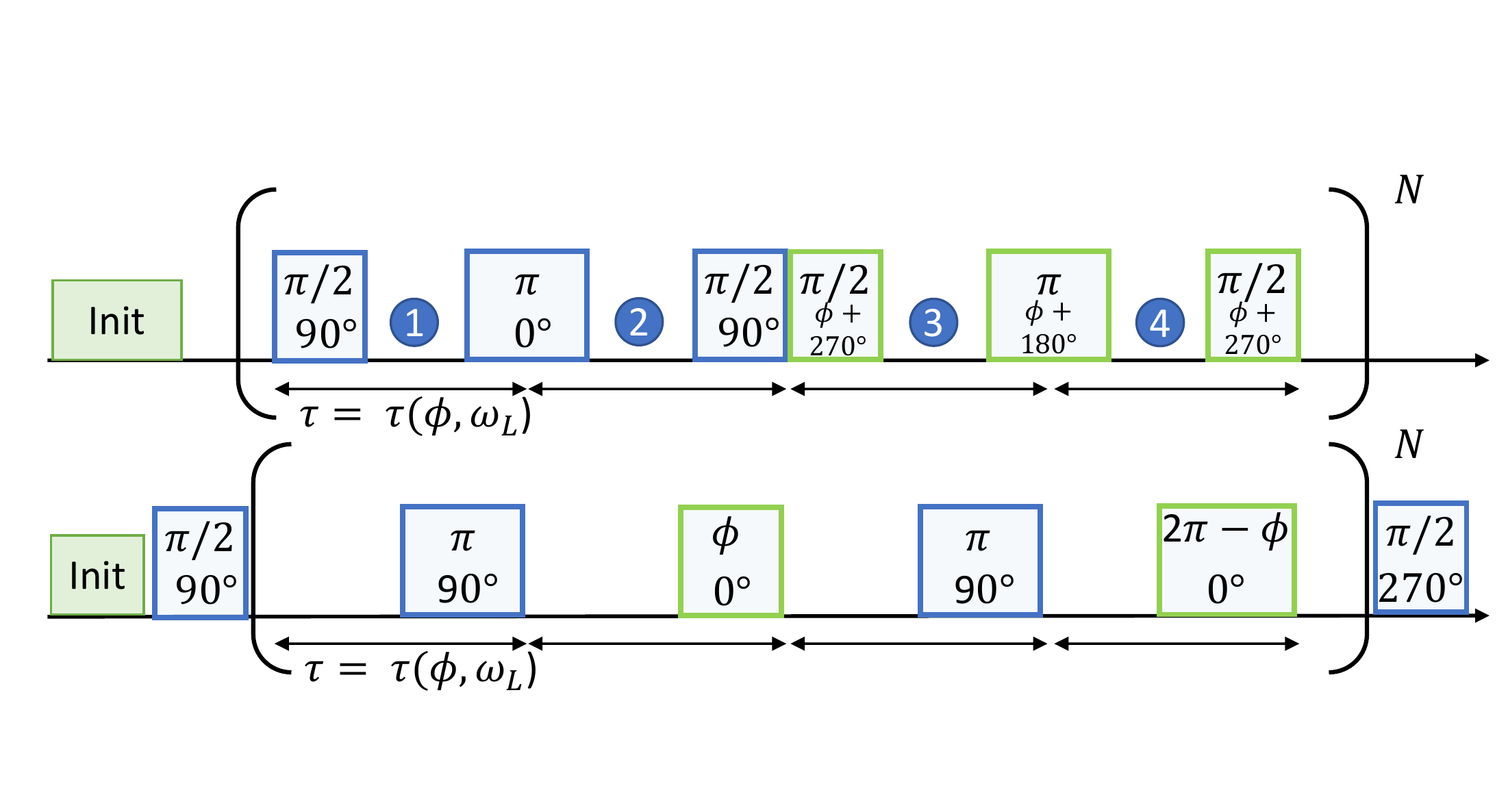}};
	\node[inner sep=0pt] (Ab) at (8.5,-1.5)
	{\includegraphics[width=.35\textwidth, trim=10cm 8cm 14cm 2cm, clip]{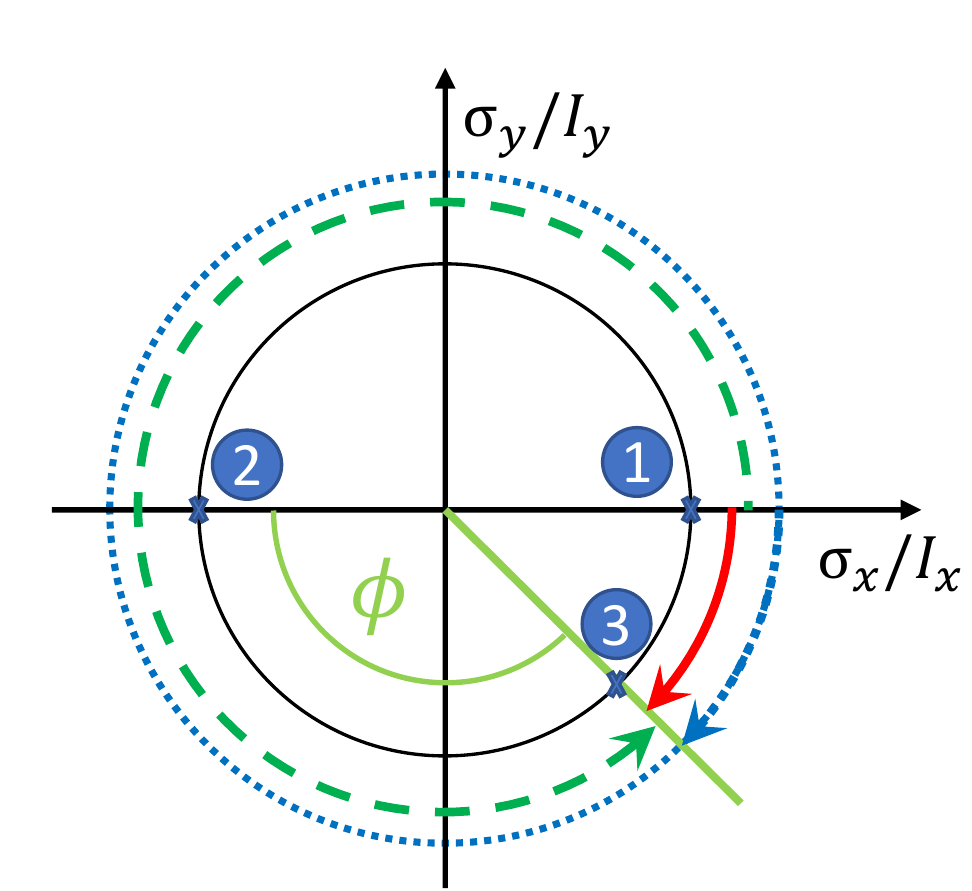}};
	\node[inner sep=0pt] (Ac) at (0,-4)
	{\includegraphics[width=.65\textwidth]{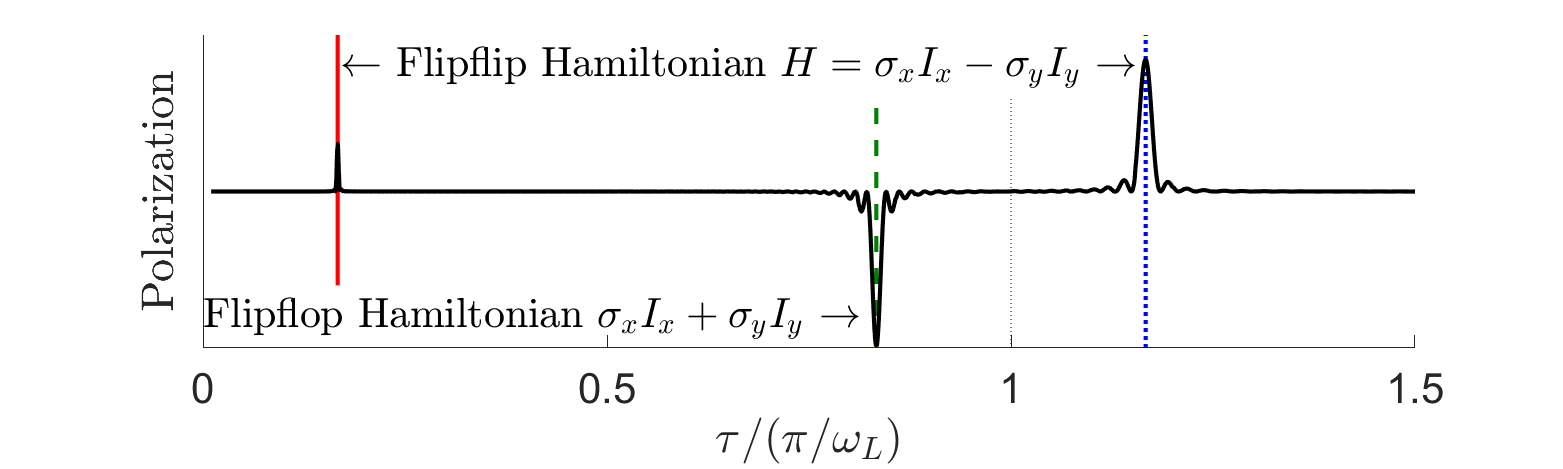}};	
	\node[inner sep=0pt] (Aa1) at (-5,1.5)
	{\Large a)};
	\node[inner sep=0pt] (Aa1) at (-5,-.5)
	{\Large b)};
	\node[inner sep=0pt] (Aa1) at (-5,-2.5)
	{\Large d)};
	\node[inner sep=0pt] (Aa1) at (6.5,1)
	{\Large c)};
	\end{tikzpicture}
	
	\caption{Two basic robust tunable pulsed polarization sequences are shown on the left top with pulse duration on the top of each pulse in radians and pulse phase at the bottom in degrees. a) PulsePol \cite{PP}, b) XY-like: the parameter $\phi$  allows to manipulate the resonances as explained in the main text, the numbers in the pulses indicate pulse length and phase, the sequence is repeated $2N$ times.
		\\
		c) The circle shows a visualization of the intuitive picture described in the text: To be resonant, the nuclear spin must cover a total angle of $\pi+\phi$ (green dashed arrow). Rotating in the opposite direction also achieves resonance (red solid arrow), as well as adding multiples of $2\pi$ (blue dotted).
		These resonances are at their expected position as illustrated in panel d), where the black line indicates the XY-resonance and the others correspond to the arrows in the right figure. If on resonance with the nuclear spin, the effective Hamiltonian transfers polarization to it, but the direction (flip-flip or flip-flop Hamiltonian $H \sim \sigma_x I_x \pm \sigma_y I_y$, corresponding to clock- or anticlockwise rotation in panel c)) depends on the resonance. }
	\label{Fig1}
\end{figure*}

Polarization sequences \cite{PP} extend this concept and create an effective $\sigma_x I_x \pm \sigma_y I_y$ interaction.
Furthermore, they can be tuned to resonance with two arbitrary frequencies $\omega_j(j=1,2)$, which can be achieved with either of the two sequences illustrated in \CHANGE{Fig.~\ref{Fig1} (a) and (b)}.
In both cases, the control over the resonance frequencies is achieved by a parameter $\phi$ that affects the pulse phases (\CHANGE{Fig.~\ref{Fig1} (a)}) or pulse duration (\CHANGE{Fig.~\ref{Fig1} (b)}).
Note that for $\phi = \pi/2$ we recover the PulsePol sequence \cite{PP} for the sequence in \CHANGE{Fig.~\ref{Fig1} (a)}, while for $\phi = \pi$ we obtain a standard XY-sequence for both sequences.
In Appendix A we analyze the robustness of both sequences against control errors, that is detuning and amplitude deviations of the driving field which are corrected by the carefully chosen pulse phases.
In particular we show the sequence in \CHANGE{Fig.~\ref{Fig1} (a)} is robust to first order errors up to a term independent of the number of sequence repetitions $N$.
Moreover, the equidistant $\pi$-pulses cancel slow dephasing noise originating from drifts of the externally applied magnetic field and magnetic impurities in the diamond sample.
For the rest of this work we will focus on the sequence a) due to its better robustness against amplitude errors but the ideas presented here apply to the sequence b) in a similar way.

In the following we want to derive the conditions under which an effective $\sigma_x I_x \pm \sigma_y I_y$ interaction is created. For this purpose we determine the modulation of the $\sigma_z$ operator that appears in \CHANGE{Eq.~}(\ref{eq4}) in the rotating frame of the control (i.e., the subsequently applied pulses). In this derivation we assume instantaneous pulses, the calculation for non-instantaneous pulses is used in Appendix C to derive the effective coupling strengths.

	The unitary evolution after the first three pulses
	can be rewritten in the rotating frame of the control as

\begin{align}
	U_1 &= e^{-i \frac{\pi}{2} \sigma_y/2}  U_\text{free}(\sigma_z, \tau)  e^{-i \pi \sigma_x/2}  U_\text{free}(\sigma_z, \tau)  e^{-i \frac{\pi}{2} \sigma_y/2} \nonumber
	\\&=
	e^{-i \frac{\pi}{2} \sigma_y/2} e^{-i \pi \sigma_x/2} e^{-i \frac{\pi}{2} \sigma_y/2}   U_\text{free}(-\sigma_x, \tau)   U_\text{free}(\sigma_x, \tau) \nonumber
	\\&=
	\pi_X U_\text{free}(-\sigma_x, \tau)   U_\text{free}(\sigma_x, \tau),
\end{align}
where $\pi_X = e^{-i \pi \sigma_x/2}$ is a $\pi$-pulse and we use the interaction Hamiltonian of \CHANGE{Eq.~}(\ref{eq6}) in the free evolution 
\begin{equation}
	U_\text{free}(\sigma_j, \tau) \equiv \exp\left(-i \tau (\omega I_z + \frac{\sigma_j}{2} (a_\perp I_x + a_\parallel I_z))\right),
\end{equation}
simplifying to only one nuclear spin for the moment.

	This means on positions 1 and 2 of the sequence in \CHANGE{Fig.~\ref{Fig1} (a)} we modulated $\sigma_{z}$ to $\sigma_x$ and $\sigma_{-x}$, which is also illustrated in \CHANGE{Fig.~\ref{Fig1} (c)}.
Continuing this scheme for the next two pulses, we make use of the phase $\phi + \pi$  that all pulses in the second part of the sequence have compared to first part. These phases can be interpreted as additional rotation around the z-axis, for example the $\pi_{\phi+180^\circ} = \pi_{\phi+\pi}$ pulse is then $R_+ e^{-i \pi \sigma_x/2} R_-$
where $R_\pm = e^{\pm i (\phi+\pi)/2 \sigma_z/2}$ describes the rotation.
As $R_\pm$ commutes with all free evolutions, we immediately see that the unitary containing the free evolutions 3 and 4 with the surrounding pulses is
\begin{equation}
	U_2 = R_+ U_1 R_-.
\end{equation}

Those unitary evolutions are repeated N times in the sequence, so we can use $R_- \pi_X = \pi_X R_+$ and the commutation of $\pi_X$ and $U_\text{free}(\pm\sigma_x, \tau)$ to express the total evolution as
\begin{align} \label{eqtot}
	U_\mathrm{tot} &= (U_2 U_1)^N
	\\& = \left( R_+ \pi_X U_\text{free}(-\sigma_x, \tau)   U_\text{free}(\sigma_x, \tau) \nonumber
	\right.\times \\& \hspace{1cm}\left.R_- \pi_X U_\text{free}(-\sigma_x, \tau)   U_\text{free}(\sigma_x, \tau) \right)^N \nonumber
	\\& = \left( R_+ U_\text{free}(-\sigma_x, \tau)   U_\text{free}(\sigma_x, \tau) \right)^{2N} \nonumber
	\\& = R_+^{2N} \prod\limits_{k=0}^{2N-1}   U_\text{free}(-R_-^k\sigma_x R_+^k, \tau)   U_\text{free}(R_-^k\sigma_x R_+^k, \tau).\nonumber
\end{align}

This shows that the rotation on all phases of the second part of the sequence $U_2$ was constructed such that the free evolution operator
$U_\text{free}(\sigma_x, \tau)$ is rotated every $2\tau$ by an additional phase $\phi+\pi$ in addition to the $\pi$ rotations between position 1 and 2, 3 and 4 etc.
A graphical representation is presented in \CHANGE{Fig.~\ref{Fig1} (c)}, where the first effective free evolution $U_\text{free}(\sigma_x, \tau)$ is depicted as position 1. The second effective free evolution $U_\text{free}(-\sigma_x, \tau)$ at position 2 is followed by $U_\text{free}(R_-\sigma_x R_+, \tau) = U_\text{free}(\sigma_x \cos(\phi+\pi) + \sigma_y \sin(\phi+\pi), \tau)$ at position 3.

This allows to rewrite the Hamiltonian in a rotating frame with respect to a nuclear Larmor frequency 
as
\begin{align}\label{eq10}
	H =   &\frac{\sigma_x \cos \phi_\text{eff}(t) + \sigma_y \sin \phi_\text{eff}(t)}{2} \\& \hspace{.2 cm}\left[ a^\perp (I_x \cos(\omega t) - I_y \sin(\omega t)) + a^\parallel I_z\right],\nonumber
\end{align}
with piecewise constant functions $\phi_\text{eff}(t)$. The interaction does only contain nonrotating terms that will not vanish during a long evolution when the function $\phi_\text{eff}(t)$ has a Fourier component with frequency $\omega$, and the $a^\parallel$ term always vanishes after the
application of the rotating wave approximation yielding the
effective Hamiltonian
\begin{equation}\label{eq10a}
	H_\text{eff} =   \frac{a^\text{eff}}{2 } \left(\sigma_x I_x \pm \sigma_y I_y \right).
\end{equation}
In the visualization of \CHANGE{Fig.~\ref{Fig1} (c)}, the nuclear spin operators of \CHANGE{Eq.~}(\ref{eq10}) proportional to $a^\perp$ evolve with a periodicity of $2\pi/\omega$. As calculated in \CHANGE{Eq.~}(\ref{eqtot}), the effective $\sigma_z$ operator gains a phase $\pi\pm\phi$ every $2\tau$, depending on if we view the rotation as being clock- or anticlockwise. Resonance is achieved if two free evolution periods of a time $2\omega\tau$ result in the same phase for the electron and nuclear spin operators up to $2\pi n$, where n is an integer.

We conclude from this picture that all nuclear spins whose Larmor frequencies $\omega$ fulfil
\begin{equation}\label{gateeqAX}
	2\tau = \left(n + \frac{1}{2} \pm \frac{\phi}{2\pi}\right) \frac{2\pi}{\omega}
\end{equation}
interact resonantly with the electronic spin, while the polarization direction depends on the $\pm$ sign ($\sigma_x$ rotating in same/opposite direction as the $I_x$ operator).
Here, $n>0$ is the the order of the resonance corresponding to the number of full cycles per step of 2$\tau$, which motivates to denote a resonance $\tau$ by (n, $\pm$) in the following.
For the specific case $\phi=\pi/2$ we recover the PulsePol resonances $\tau \omega/\pi = 1/4, 3/4, 5/4, ...$ \cite{PP}.
\CHANGE{Fig.~\ref{Fig1} (d)} shows the first three resonances for $\phi = 3\pi/4$.

In order to simultaneously put two chosen frequencies into resonance, the free parameter $\phi$ may be chosen as follows.
For example, for frequencies $\omega_1 < \omega_2 < 3 \omega_1$ we can use the $(n=0,+)$ and $(n=1,-)$ conditions.
Hence we need to solve

\begin{equation}
	\left(\frac{1}{2} + \frac{\phi}{2\pi}\right) \frac{2\pi}{\omega_1} = \left(\frac{3}{2} - \frac{\phi}{2\pi}\right) \frac{2\pi}{\omega_2}
\end{equation}

for $\phi$, which yields

\begin{equation}
	\phi = \frac{\pi}{2} \frac{3-\omega_2/\omega_1}{1+\omega_2/\omega_1}.  \label{polcond1}
\end{equation}

Here, the nuclear spins precessing at two frequencies $\omega_1$ and $\omega_2$ 
are polarized in opposite directions.
Polarization in the same direction, i.e. applying a flipflop-interaction for both nuclei instead of applying  a flipflop-interaction to one nucleus and a flipflip-interaction to the other, can be achieved by employing the $(n,\pm)$ and $(n+1,\pm)$ conditions, where we obtain
\begin{equation}
\pm \phi = 2\pi \frac{1}{\omega_2/\omega_1-1} - n-\frac{1}{2}.
\end{equation}

We remark that the effective electron-nuclear coupling strength varies depending on the chosen resonance $(n, \pm)$ and the parallel coupling $a_\perp$.
In the case of instantaneous pulses we obtain the effective coupling
\CHANGE{
\begin{align}\label{eq15}
a^\text{eff} 
%&= \frac{a^\perp}{2n\pi+\pi\pm\phi}
%\left( 1  + \sin \left(n\pi \pm \frac{\phi}{2}    \right)                \right)
=\frac{a^\perp}{2\tau\omega_L} \left[1-\cos(\omega_L \tau)  \right]
\end{align}
for the $(n, \pm)$ resonance where $\omega_L \tau /\pi = n + 1/2 \pm \phi/(2\pi)$ according to \CHANGE{Eq.~}\eqref{gateeqAX}.
}

This value is constant once the resonance $(n, \pm)$ is fixed. However, for applying these sequences in quantum information processing or simulation, it is required to tune the effective coupling strength for different resonances independently, otherwise only one specific interaction that depends on the system configuration can be created.
In the next section we
solve this limitation by extending the sequences to allow for an arbitrary tuning of the coupling strength.

\section{III. Using Polarization sequences for high-fidelity nuclear-nuclear gates}

We will now show how we can tune the coupling strengths $a_1$ and $a_2$ to two nuclear spins which we aim to couple simultaneously to the NV center. This allows to realize arbitrary entangling gates between nuclei with different coupling strength and Larmor frequency.
There are two general gates to distinguish, that is, whether
the nuclei interact with the electron spin with one flipflip and one flipflop interaction or two flipflop (flipflip) interactions.
\CHANGE{While the gate operators differ for these cases, the distinction is only relevant in the case of similar Larmor frequencies $|\omega_2/\omega_1-1| \ll 1$ that is discussed in the following section.} 

Considering the second case, the effective evolution operator under the developed sequences for polarization in the same direction might be written as
\begin{align}
	U(t)=\exp & \left\lbrace-i t/2 \left[ a_1(\sigma_x I_x^{1} + \sigma_y I_y^{1}) \right.\right. \nonumber \\
	&\quad\quad\left.\left.+ a_2(\sigma_x I_x^{2} + \sigma_y I_y^{2})    \right]  \right\rbrace.
\end{align}
For an equal coupling strength $a_1=a_2$ and an evolution time $t$ such that $\sin( \sqrt{a_1^2+a_2^2}t/4) = 1$, the operator simplifies to (see Appendix C)
\begin{align}
	V =&\mathbb{1} \left(- \ket{10}\bra{01} -
	\ket{01}\bra{10}\right)\nonumber
	\\&+
	\sigma_{z}  \left(\ket{11}\bra{11} - \ket{00}\bra{00}\right).
\end{align}
Within the nuclear spin subspace, this gates acts like the composition of a \mbox{CPHASE} and a  \mbox{SWAP} gate.
Together with local rotations on the nuclear spins, this would be sufficient for universal quantum computation \cite{nielsen2002quantum}.
For example, application of this gate in combination with a $\pi/2$ pulse around the $y$ axis on the second nuclei can create the minimal GHZ state
\begin{align*}
 e^{-i\pi/2 I_y^{(2)}}V|0\,y_-\,y_-\rangle = |0\rangle \frac{|00\rangle+|11\rangle}{\sqrt{2}}.
\end{align*}
%.
These states are maximally entangled and advantageous in quantum metrology \cite{HuelgaMP+97}.
\\

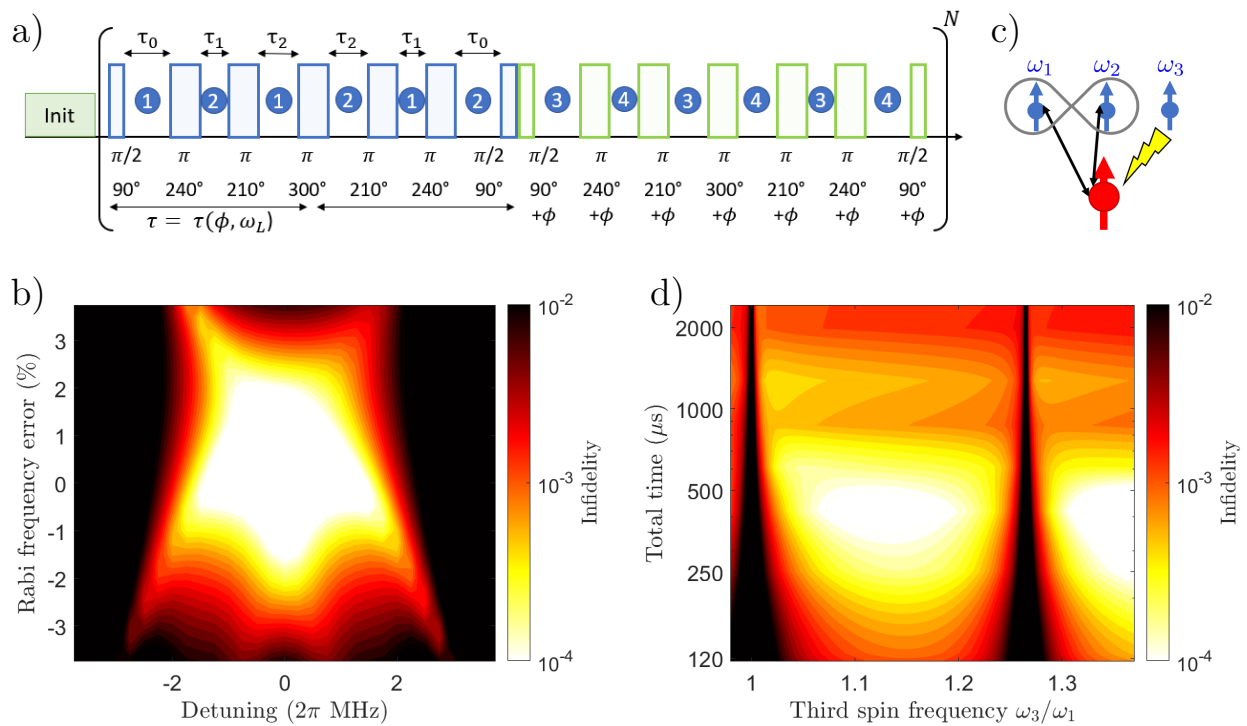
\begin{figure*}[ht]
	\centering
	\begin{tikzpicture}
		\node[inner sep=0pt] (A) at (6-3.5,5)
		{\includegraphics[width=.7\linewidth]{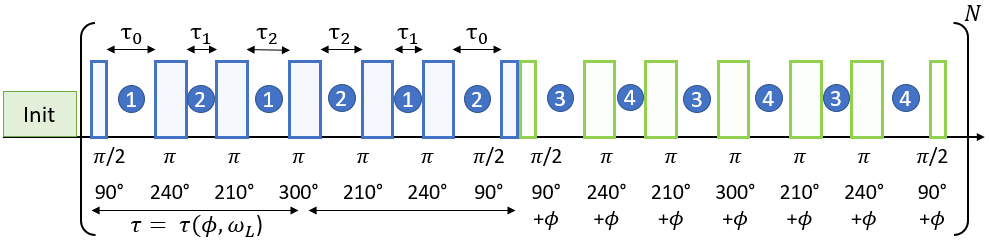}};
		\node[inner sep=0pt] (Ac) at (-2+12.5,4.75)
		{\includegraphics[width=.15\linewidth]{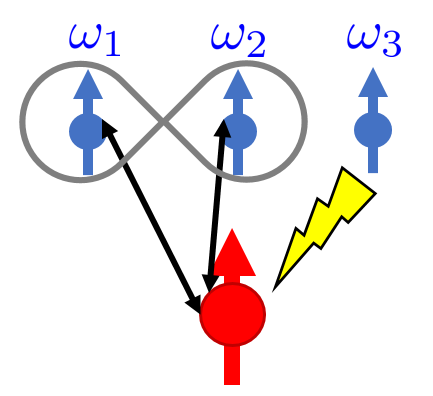}};
		\node[inner sep=0pt] (Aa) at (0,0)
		{\includegraphics[width=.45\linewidth]{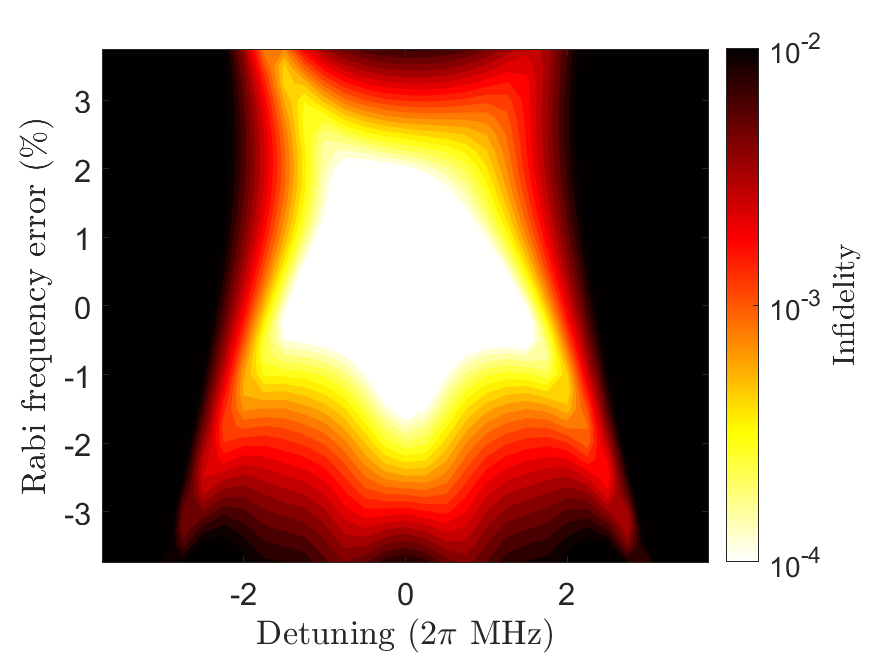}};
		\node[inner sep=0pt] (Ab) at (8.5,0)
		{	\includegraphics[width=.45\linewidth]{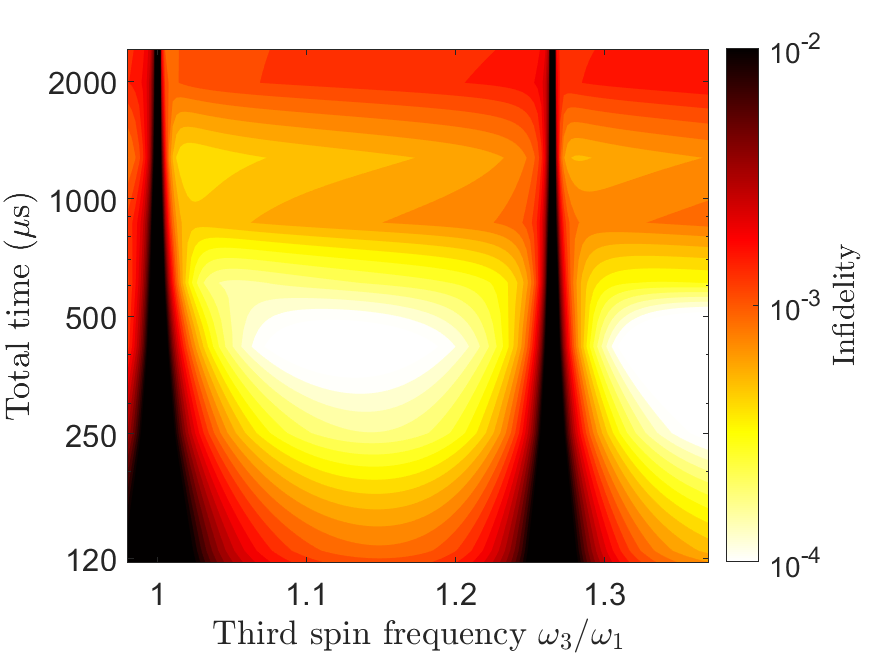}};
		
		\node[inner sep=0pt] (Aa1) at (-3.7,2.5+.4-.2)
		{\Large b)};
		\node[inner sep=0pt] (Aa1) at (-3.7,6.2)
		{\Large a)};
		\node[inner sep=0pt] (Aa1) at (-3.7+8.5,2.9-.2)
		{\Large d)};
		\node[inner sep=0pt] (Aa1) at (-3.7+13,6.2)
		{\Large c)};
	\end{tikzpicture}
	\caption{
		a) Sequence used for all following simulations: The $\pi$ pulses in the sequence of \CHANGE{Fig.~\ref{Fig1} (a)} are replaced by 5 pi-pulses, introducing the parameters $\tau_{0,1,2}$.
		b) Process infidelity for different detuning and amplitude errors using the (0,+) and (1,-) conditions, for details see text.
		c) Schematic visualization of the situation considered: An NV center is used to entangle 2 nuclear spins of frequencies $\omega_{1/2}$, while a third spin of frequency $\omega_3$ inhibits the gate.
		d) Infidelity as a function of the allowed total gate time and the frequency of an additional spin.			
	}
	\label{Fig2}
\end{figure*}

Similar results are obtained using the gate 
implemented with one flipflop and one flipflip interaction,
that is
\begin{align}
U(t)=\exp & \left\lbrace-i t/2 \left[ a_1(\sigma_x I_x^{1} + \sigma_y I_y^{1}) \right.\right. \nonumber \\
&\quad\quad\left.\left.+ a_2(\sigma_x I_x^{2} - \sigma_y I_y^{2})    \right]  \right\rbrace.
\end{align}
In particular, for equal coupling and the gate time fulfilling $\sin( \sqrt{a_1^2+a_2^2}t/4) = 1$ we arrive at
\begin{align}
V = &\mathbb{1} \left(-\ket{11}\bra{00} - \ket{00}\bra{11}\right)\nonumber
\\&+
\sigma_z \left(\ket{10}\bra{10} -
\ket{01}\bra{01} \right).
\end{align}

However, the condition $a_1=a_2$ only holds for the special case where the ratio between the Larmor frequencies is the ratio between the perpendicular coupling strengths and a miss-match heavily decreases the fidelity of the gate.
We thus extend the PulsePol sequence in \CHANGE{Fig.~\ref{Fig1} (a)} by replacing each $\pi$-pulse by a composite pulse as proposed in \cite{tycko1984iterative, ryan2010robust, souza2011robust, AXY}.
%As explained earlier around Hamiltonian (11) the application of pulse sequences map the \sigma_z operator in the interaction term of Hamiltonian (4) to a time-dependent effective operator of the form \sigma_x  cos(\phi_eff t) + \sigma_y sin(\phi_eff t). 
%
\CHANGE{A $\pi$-pulse interchanges the populations in $\ket{0}$ and $\ket{1}$, which is equivalent to the transformation $\sigma_z\mapsto -\sigma_z$. 
Hence, we can model these pulses by a modulation function $f_1(t)=(-1)^{n(t)}$, where $n(t)$ is the number of $\pi$-pulses applied until $t$, i.e. it is positive (negative) after an even (odd) number of pulses.  
Combining this observation with the derivation of the effective Hamiltonian in Eq.~\eqref{eq10}, we now have that the total effective coupling operator of the NV center reads $f_1(t)[\sigma_x  \cos(\phi_\mathrm{eff} t) + \sigma_y \sin(\phi_\mathrm{eff} t)]$.
If the value of $\phi$ in the pulse sequence [see Fig.~\ref{Fig1}(a)], equals $\pi$, we have that $f_1(t)=\cos(\phi_\mathrm{eff} t)$ and $ \sin(\phi_\mathrm{eff} t) =0$, i.e. we obtain the effective operator of the XY-4 sequence. Otherwise, the function $\phi_\text{eff}(t)$ differs by the periodical phase changes $\phi+\pi$ that are introduced every $2\tau$ as described in section II.}
%\CHANGE{\CHANGEB{As explained earlier around Hamiltonian (\ref{eq10}) the application of pulse sequences map the $\sigma_z$ in the interaction term of Hamiltonian (\ref{eq4}) to a time-dependent effective operator of the form $\sigma_x  \cos(\phi_\mathrm{eff} t) + \sigma_y \sin(\phi_\mathrm{eff} t)$.}
%The additional pulses change the sign of this operator, which can be described by a periodic modulation function $f_1(t)$ that equals $\pm1$ after an even (odd) number of $\pi-$pulses.
%	\CHANGEB{For $\phi=\pi$, i.e. when the second block of the pulse sequence is identical to the first block and a XY-4 sequence is obtained, this function equals the term $\cos \phi_\text{eff}(t)$ in Eq.~(\ref{eq10}), otherwise the function $\phi_\text{eff}(t)$ differs by the periodical phase changes $\phi+\pi$ that are introduced every $2\tau$ as described in section II.}
\CHANGE{As shown in \CHANGE{Fig.~\ref{Fig2} (a)}, two new parameters $\tau_1$ and $\tau_2$ that describe the new pulse spacings are introduced and can be employed to tune the Fourier components of the modulation function $f_1(t)$.	
	As a result, the effective coupling is individually modified for each resonance because the relevant Fourier components can be adjusted independently, see appendix C for details.}
%As shown in Figure \ref{Fig2} a), this introduces two new parameters $\tau_1$ and $\tau_2$ that can be employed to tune the effective coupling to the nuclei.
In particular, for instantaneous pulses it reads
\CHANGE{
\begin{align}
a_\text{eff} &=  
%\frac{a^\perp}{\pi+\phi_{n, \pm}}
%\left[ 1  + 2\cos \frac{(\beta_1+\beta_2)(\phi_{n, \pm} + \pi)}{2}
%\right.\\&\hspace{1 cm}\left.
%- 2\cos \frac{\beta_2(\phi_{n, \pm} + \pi) }{2}    + \sin \frac{\phi_{n, \pm} }{2}                    \right],
\frac{a^\perp}{2\tau\omega_L} 
\\&\left[1-\cos(\omega_L \tau) -2\cos(\omega_L \tau_2) +2\cos(\omega_L (\tau_1+\tau_2)) \right].\nonumber
\end{align}
%
%where we defined %$\tau_{1,2} = \beta_{1,2} \frac{\pi+\phi}{2\omega_L}$
%$\beta_{1,2} = \tau_{1,2} \frac{2\omega_L}{\pi+\phi}$ 
%and $\phi_{n, \pm} = 2n\pi \pm \phi$.
Here, unlike in \CHANGE{Eq.~}(\ref{eq15}), the parameters $\tau_{1,2}$ can be used to modify the effective coupling of a given resonance.
} 
Using this expression
the right values of the $\tau_j$ for a given pair of nuclear spins and the desired effective coupling strengths $a_i^\perp$ can be deduced numerically. The expression for non-instantaneous pulses is given in Appendix C.

We apply this concept and evaluate the performance of the gate under amplitude and detuning errors of the drive which induces the pulses.
To do so, we employ the process fidelity of the gate, which is given by the overlap of the Choi states corresponding to the implemented evolution via the pulse sequences and the desired target evolution \cite{altepeter2003ancilla}.
\CHANGEB{This translates to calculating the fidelity as the overlap of the states $\Psi(U) = U \otimes \mathbb{1} \sum\limits_{i=0}^{d-1} \ket{i}\otimes \ket{i} /\sqrt{d}$, where $U$ is either the unitary evolution under the described gate or the desired target evolution, while $\ket{i}$ are the basis states of the NV-nuclear system with dimension $d=2^4$ for one electron spin and three nuclear spins.}

%This means calculating the fidelity of the states $\rho_\rm{Choi}(U) = the formula$, where $U$ is either the unitary evolution under the described gate or the desired target evolution, while $\ket{i}$ ...
%\CHANGE{This metric allows to characterize the fidelity of an operation for all initial states by simulating the unitary evolution of a system that is maximally entangled with an ancillary system and therefore allows to capture gate infidelities for specific linear combinations of initial states that can be overlooked when only the fidelity for a set of initial states is considered.}
\CHANGE{This metric is able to characterize the fidelity of an operation concurrently for all initial states. In particular, it therefore allows to capture gate infidelities for specific combinations of initial states that might be overlooked when only the state fidelity for certain intial/target state combination is considered.} 
\CHANGE{Fig.~\ref{Fig2} (b)} shows $>$99.99\% process-fidelity for a gate with a silicon-29  ($\omega_1 = (2\pi) $2MHz) and a carbon-13 spin ($\omega_2 = 1.265\omega_1$) coupled to an NV center ($a_1^{\perp} = (2\pi) $20kHz, $a_2^{\perp} = (2\pi) $25kHz), using a Rabi frequency $\Omega = (2\pi) $50MHz and a total time $T \cong 418 \mu s$.
Pulse control errors of $<2$\% amplitude and $< (2\pi)$ 2MHz detuning errors do not affect the fidelity.
While the shape of the optimal region looks different for different gate times, this robustness is always sufficient for common experimental systems that operate with a single NV center.

Adding a third spin with $a_3^{\perp} = (2\pi) $20kHz and a frequency close to that of one of the other nuclei as sketched in \CHANGE{Fig.~\ref{Fig2} (c)} impairs 
the fidelity in a frequency range $\sim 1/T$, where $T$ is the total gate duration.
This is seen in \CHANGE{Fig.~\ref{Fig2} (d)}, where, for fixed total gate time, an infidelity larger than 1\% appears in a larger range around the first two nuclear spin frequencies.
		This is a result of the frequency filter scaling $\sim 1/T$, that stems from the effective Fourier transformation we perform in \CHANGE{Eq.~}(\ref{eq10a}).
		We investigate this in more detail after the following section.
		Note that the values for larger total gate times (decreasing fidelity for $T >700 \mu s$) can be improved by choosing a higher order resonance.

While this section used a gate with one flipflop and one flipflip Hamiltonian, we will demonstrate the application of a gate with two flipflop Hamiltonians in the example presented in the next section.

\section{IV. Manipulation of noise-protected states}

The states in the subspace \{$\ket{10}, \ket{01}$\} for nuclei with the same gyromagnetic ratio are protected from fluctuations of the global magnetic field.
Therefore, it is an excellent candidate to store quantum information since its coherence time is substantially prolonged
\cite{perlin2018noise}.
Efficient manipulation in this submanifold can further be used to store and manipulate quantum information in the protected manifold or to obtain information on the frequency difference and therefore the magnetic field gradient on a small lengthscale.

In this section we show fast high-fidelity manipulation in this submanifold using the protocol described in the previous subsection, but generalizing the operator $V$ from the previous section.
A major drawback of using the subspace \{$\ket{10}, \ket{01}$\} is that resonance conditions that polarize in the same direction, e.g. two flip-flip Hamiltonians, have to be used. \CHANGE{For similar Larmor frequencies $|\omega_2/\omega_1-1| \ll 1$, this} demands higher resonance conditions and therefore longer gate times, because we now need to solve

\begin{equation}
\left(\frac{2n+1}{2} + \frac{\phi}{2\pi}\right) \frac{2\pi}{\omega_1} = \left(\frac{2n+3}{2} + \frac{\phi}{2\pi}\right) \frac{2\pi}{\omega_2}
\end{equation}
for $\phi$ and $n$. We are particularly interested in similar frequencies where
\begin{equation}
   1 \ll \frac{\omega_2}{\omega_1} - 1 =    \frac{1}{\frac{2n+1}{2}+\frac{\phi}{2\pi}}
\end{equation}
and therefore large $n$ is required as $\phi \in \left[0, 2\pi\right]$.

In the previous section we effectively used the \{$\ket{11}, \ket{00}$\} manifold using one flip-flip and one flip-flop Hamiltonian, which is beneficial when the gates are note required to affect the \{$\ket{10}, \ket{01}$\} subspace.

\begin{figure}[t]
	\centering
		\includegraphics[width=.9\linewidth]{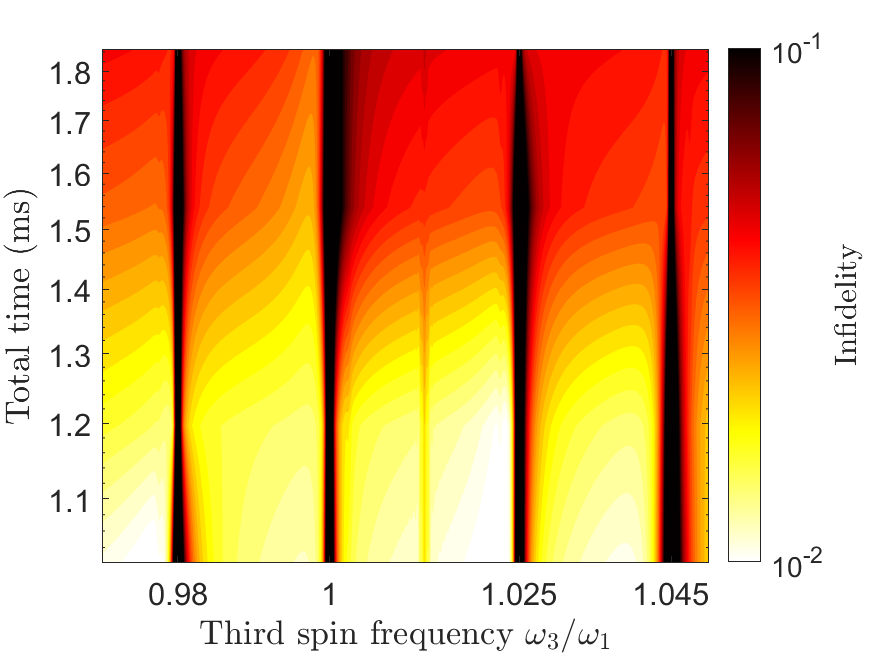}
	\caption{Process infidelity in the relevant nuclear subspace \{$\ket{10}, \ket{01}$\} of a gate that evolves a state $\ket{1} \otimes \ket{1 0} \otimes \ket{+} $ to $ \ket{1} \otimes \left( \ket{1 0} + \ket{ 01}  \right)/\sqrt{2} \otimes \ket{+} $ while being
			disturbed by a third spin like in \CHANGE{Fig.~\ref{Fig2} (d)}.
					Here the conditions (22,-) and (23,-) were used, as a result the maximally achievable coupling strength is reduced and therefore the gate times are longer.
	}
	\label{Fig3}
\end{figure}

To create the desired gate we again fix the gate time to $\sin( \sqrt{a_1^2+a_2^2}t/4) = 1$. By defining
$\cos \alpha \equiv (a_1^2-a_2^2)/(a_1^2+a_2^2)$ we obtain an evolution operator (see Appendix B for frame linestyle explanation)

\begin{align}
\tiny
U=\left(
\begin{array}{cccc | cccc}
1 & 0 & 0 &0 &0 &0 &0 &0
\\
0 & \xa{\cos\alpha} & \xb{-\sin\alpha} & 0 &
\xc{0} &0 &0 &0
\\
0 & \xb{-\sin\alpha} & \xa{-\cos\alpha}  & 0 &
\xc{0} &0 &0 &0
\\
0 & 0 & 0 &\xa{-1} &0 &\xc{0} &\xc{0} &0
\\
\hline
\\
0 &\xc{0} &\xc{0} &0 & \xa{-1} &0 & 0 & 0
\\
0 & 0 & 0  & \xc{0} &
0 &\xa{-\cos\alpha} &\xb{-\sin\alpha} &0
\\
0 & 0 & 0 & \xc{0} &
0 &\xb{-\sin\alpha} & \xa{\cos\alpha} &0
\\
0 & 0 & 0 &0 &0 &0 &0 &1
\end{array}
\right),\nonumber
\end{align}

where the space spanned by $\ket{10}$ and $\ket{01}$ is mapped onto itself, which means arbitrary rotations controlled by the NV center are performed using the unitary
\begin{align}
U = -\sin(\alpha) \mathbb{1}^\text{NV} \otimes \sigma_x^{(1)} + \cos(\alpha) \sigma_z^\text{NV} \otimes \sigma_z^{(1)},
\end{align}
where $\sigma_x^{(1)} = \ket{10}\bra{01}+ \ket{01}\bra{10}$ and
$\sigma_z^{(1)} = \ket{10}\bra{10} - \ket{01}\bra{01}$.
After initializing the nuclei into $\ket{10}$, for example by using a polarization sequence repeatedly, information can be swapped from the electron spin onto the nuclear spins or to perform local gates in the submanifold \{$\ket{10}, \ket{01}$\}. As an example, we consider a Hadamard gate on the encoded subspace in \CHANGE{Fig.} \ref{Fig3} that transforms an initially prepared state $\ket{10}$ into $(\ket{10} + \ket{01})/\sqrt{2}$. From the condition $\cos \alpha = 1/\sqrt{2}$ we obtain $a = (\sqrt{2}\pm 1)b$, which we can again control by choosing the waiting times between the 5 $\pi$-pulses correctly.

\CHANGE{Fig.} \ref{Fig3} uses the resonances (22,-) and (23,-) 				
for two nuclei ($\omega_1 = (2\pi) $2MHz, $\omega_2 = 1.045\omega_1$, $a_1^\perp = (2\pi) $20kHz, $a_2^\perp = (2\pi) $25kHz and a Rabi frequency $\Omega = (2\pi) $30MHz. An additional third spin $a_3^\perp = (2\pi) $20kHz decreases the fidelity significantly if it is close in frequency to the first or second spin or to other resonances (e.g. (21,+) at $\sim 0.98\omega_1$ and (22,+) at $\sim 1.025\omega_1$).
For higher effective couplings the fidelity is higher as less perturbations can accumulate during the smaller gate time, but also the range where additional frequencies can perturb the gate increases with decreasing gate time.

Analogously to this section it is possible to manipulate the states $\ket{11}$ and $\ket{00}$ for enhanced sensing using the approach from the previous section with one flipflip- and one flipflop-interaction but with a more general operator $V$ that is given in the Appendix B or obtained by swapping one nuclear spin basis $\ket{0} \leftrightarrow \ket{1}$ compared to this section.
Those states accumulate the two times the phase from an external magnetic field compared to a single nucleus, allowing for a more precise measurement \cite{HuelgaMP+97} due to the use of entanglement.

\section{V. Comparison to gates composed of electron-nuclear gates}

The action of the gates constructed in this work can be equivalently realized by a synthesis of electron-nuclear gates 
that achieve $\sigma_z I_x$ interactions using standard sensing sequences \cite{ZyJg}. Importantly, these gates 
are mainly limited by the decoherence time $T_2$ of the electron spin, during which at least two electron-nuclear gate operations 
have to be performed. While our work is motivated by NV centers in diamond controlling nearby C-13 spins, the same considerations also apply for shallow 
NV centers near the diamond surface that suffer from smaller decoherence times due to various defects on the diamond surface. 
However they can be used to control nuclear spins of various materials near the surface, including hexagonal boron nitride 
(hBN), silicon carbide and fluorine attached to the diamond surface. While still being prone to electronic dephasing during 
the gate time, the smaller total gate time achieved by the polarization sequence approach presented in this work can increase fidelities compared to gates synthesized by electron-nuclear 
gates.

\begin{figure}[t]
	\centering
	\includegraphics[width=.9\linewidth]{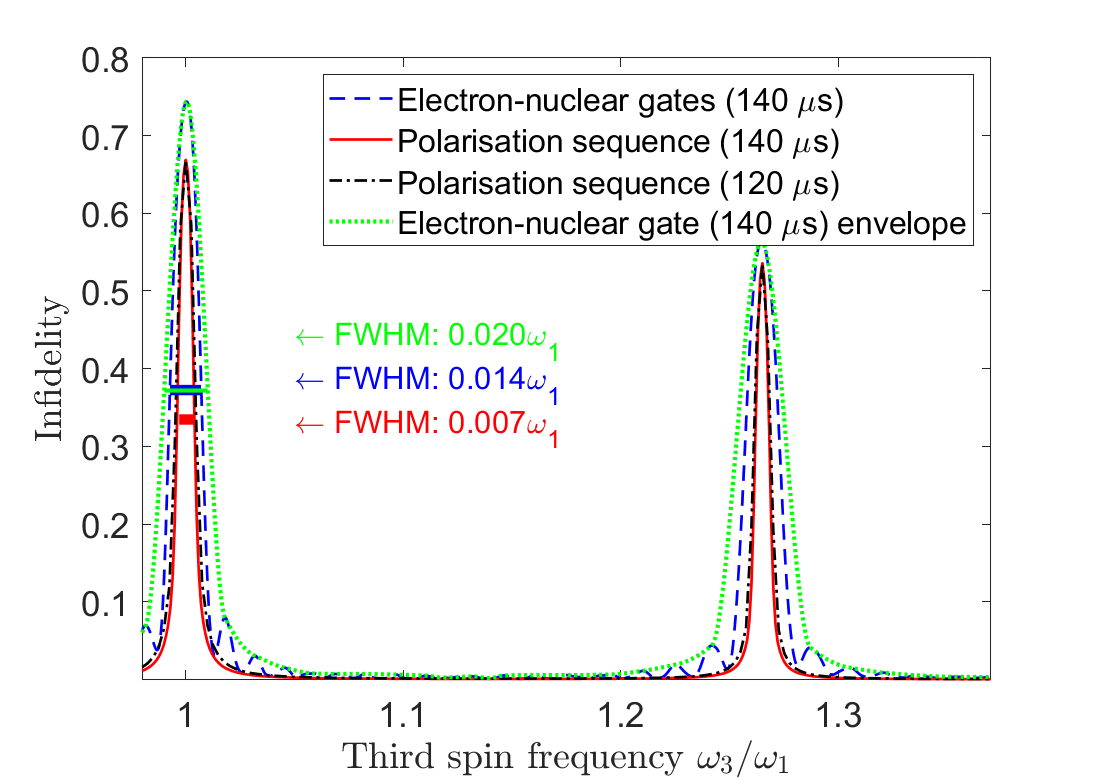}
	
	\caption{
		Comparison of process infidelities of the direct gate approach (present work) and the sequential gate approach \cite{ZyJg}. 
		For the parameters of \CHANGE{Fig.} \ref{Fig2} our proposed direct gate approach is shown for the fastest possible time $T\cong120\mu s$ (black dash-dotted line) and $T\cong140\mu s$ (red solid line), which is the fastest possible gate time achievable with a sequential gate approach. The latter is simulated using the ideal case of instantaneous-pulse CPMG sequences (blue dashed line and its envelope in green dotted line).
		The regions where a third nucleus affects the gate are smaller for the polarization based sequences. The full width half maximum (FWHM) for equal total gate times that characterizes the range of frequencies that impair the gate fidelity is almost three times worse for the electron-nuclear gate envelope compared to the polarisation sequence approach. A factor 2 that stems from the fact that electron-nuclear gates only use half of the sequence time to decouple the individual nuclei can be seen in the blue curve FWHM, its sinc-shape is the reason for the bigger envelope, see text for discussion.
	}
	\label{Fig4}
\end{figure}

Let us consider two limiting factors. The first one applies when operating in samples where nuclear spins are abundant. In this 
case we are faced with a complicated and dense environmental frequency spectrum which requires a narrow frequency window 
to ensure selective addressing of nuclei. 
\CHANGE{Fig.} \ref{Fig4} shows a comparison of our direct gate approach to a gate composed of electron-nuclear gates as proposed in \cite{ZyJg}.
For this example polarization based sequences are not only faster (see second case), but also provide better robustness to disturbance from a third spin.
For equal total gate times, the frequency range where a third spin disturbs the nuclear gate (red solid curve) is only about one third of the range for the electron sequence envelope counterpart (green dotted curve).
Our approach employing polarization sequences then only requires  50\% of the electron-nuclear 
composed gate time since at all times two nuclei interact with the electron spin. This is reflected in the blue curve FWHM and clear from the fact that the nuclear-nuclear 
gate approach manipulates both frequencies simultaneously, while the electron-nuclear gates address the two frequencies during two different 
parts of the sequence. Note that for electron-nuclear gates only half of the total sequence is used to decouple the respective frequency 
from the environment, since two gates have to be applied. The property of decoupling both nuclei simultaneously while entangling them 
at the same time is fundamentally related to the fact that for polarization sequences the effective interaction operators for the respective 
nuclei, i.e. $\sigma_x I_x^{1}+\sigma_y I_y^{1}$ and $\sigma_x I_x^{2}\pm \sigma_y I_y^{2}$, do not commute. Contrary, the effective 
operators $\sigma_x I_x^{1}$ and $\sigma_x I_x^{2}$ created via the sensing sequences commute and therefore cannot be used to entangle simultaneously.
Another advantage of polarization sequence based gates in \CHANGE{Fig.} \ref{Fig4} is that there are no sidepeaks of infidelity, unlike for the electron-nuclear gate approach. 
	In the latter case different sinus cardinalis (sinc) expressions appear in the frequency domain filter \cite{ajoy2011optimal, haase2018soft}, resulting in a broader envelope.
	In typical experimental settings, where a given fidelity needs to be achieved by decoupling from an impairing spin bath with a range of frequencies, it is required to compare the envelopes of the curves. The envelope for the electron-nuclear gate approach (green dotted) shows an almost three times larger FWHM than the polarization sequence based approach for the same total gate time.
\\

In cases where frequency selective addressing is not the limiting factor, the strength of the available effective dipole-dipole coupling
$A_\text{eff}^\perp$ between the electron and nuclear spins becomes a key factor as it determines the gate time for a given gate angle. In the
ideal case where the coupling strengths already have the preferred ratio, e.g. $\omega_1 / \omega_2 = a_1^\perp/a_2^\perp = 5/3$ for $\phi = 
\pi/2$ and the $(1,-)$ condition, we use \CHANGE{Eq.~}(\ref{eq15}) and $\sin(4a\sqrt{2}t)=1$ to obtain the total gate time $T = 3 \pi^2 /(1+\sqrt{2}) 
/ a_1^\perp$, whereas an approach via standard sensing sequences \cite{ZyJg} need two blocks of time $\pi/(2a_{1,2}^\perp)$ for each nucleus, 
so a total time of T = $8\pi^2/(5 a_1^\perp)$. Therefore the polarization approach needs up to 22\% less time. In a realistic scenario 
with finite pulses and non-ideal coupling ratios this improvement is slightly smaller, yielding a reduction in gate time of $14\%$ for the 
parameters in \CHANGE{Fig.} \ref{Fig4}.

Summarizing, our approach can reduce the total gate time unless a gate between spins with coupling strengths that differ by orders of magnitude 
is required, which allows for a higher number of operations within the electron $T_2$ coherence time, representing the natural limit.

\section{VI. Conclusion}

In this work we presented an extension to pulsed polarization sequences that, in contrast to standard techniques, allows to address simultaneously 
two nuclear spin species with distinct frequencies of similar magnitude. We show that the resulting effective flipflop and flipflip electron-nuclear interactions can be used 
to realize entangling nuclear-nuclear gates and describe how to manipulate the effective Hamiltonian to perform desired gates by introducing additional pulses. Finally we show 
that these gates are faster than gates composed of single electron-nuclear interactions, in particular when simultaneous decoupling from other spins is required, and discuss 
relevant experimental settings.

\section{Acknowledgements}
%This work was supported by the ERC
%Synergy Grants BioQ (Grant no 319130) and HyperQ (Grant no 856432), the EU projects AsteriQs and HYPERDIAMOND, the BMBF projects NanoSpin and DiaPol, 
%the IQST and the DFG via a Reinhart Koselleck project.
%JFH acknowledges the Alexander von Humboldt Foundation in the form of a Feodor-Lynen Fellowship. 
%\CHANGE{ZYW acknowledges support by the National Natural Science Foundation of China (Grants No. 12074131).}
%The authors acknowledge support by the state of
%Baden-W{\"u}rttemberg through bwHPC and the German Research Foundation (DFG) through grant no
%INST 40/467-1 FUGG (JUSTUS cluster).
This work was supported by the ERC Synergy Grants BioQ (Grant No. 319130) and HyperQ (Grant No. 856432), the EU projects AsteriQs (Grant No. 820394) and HYPERDIAMOND (Grant No. 667192), the BMBF projects NanoSpin, DiaPol and Q.Link.X, the IQST and the DFG via a Reinhart Koselleck project. J.F.H. acknowledges the Alexander von Humboldt Foundation in the form of a Feodor-Lynen Fellowship. Z.Y.W. acknowledges support by the National Natural Science Foundation of China (Grants No. 12074131). The authors acknowledge support by the state of Baden-Württemberg through bwHPC and the German Research Foundation (DFG) through Grant No. INST 40/467-1 FUGG (JUSTUS cluster).

\bibliographystyle{MyUnsrtnat}
\bibliography{bibPPX}

\newpage
\onecolumngrid
\section{Appendix}

\subsection{Appendix A: Sequence robustness to errors}

We can describe every pulse with a unitary evolution
	\begin{equation}
		U = \exp \left(-i \Omega_r t \left(\delta \sigma_z/2 + (1+\epsilon) (\cos \varphi \sigma_x/2 + \cos \varphi \sigma_y/2)\right)\right)
	\end{equation}
where $\Omega_r t$ is the intended rotation angle (e.g. $\pi$ for a $\pi$-pulse) with the pulse duration $t$ and the intended Rabi frequency $\Omega_r$, $\delta = \Delta/\Omega_r$ describes detuning errors and $\epsilon = \Omega/\Omega_r-1$ describes driving amplitude errors.
Analyzing the pulse sequence in \CHANGE{Fig.~\ref{Fig1} (a)}, we see that Rabi frequency errors are cancelled to first order and the detuning errors lead in first order to a correction term after $2N$ repetitions
\begin{equation}
		U_\text{sequence, 2N} = U_\text{perfect-pulses} \times (1+\delta \begin{pmatrix}
		0 & (1+e^{i\phi})(1+e^{2 i\phi})...(1+e^{2Ni\phi}) \\ (1+e^{-i\phi})(1+e^{-2 i\phi})...(1+e^{-2Ni\phi}) & 0
		\end{pmatrix}).
\end{equation}

The magnitude of the correction term is bounded (unless $\phi=0+2k\pi$ which is not useful for our sequence even without errors) as we can calculate the offdiagonal terms using a gemoetric series

\begin{equation}
	\left|(1+e^{i\phi})(1+e^{2 i\phi})...(1+e^{2Ni\phi})\right| = \left|\frac{1-e^{4Ni\phi}}{1-e^{i\phi}}\right| \le 1/\cos(\phi/2))
\end{equation}

In particular, for $\phi=\pi/2$ (PulsePol \cite{PP}) first order errors cancel already for $N=1$, and the XY4-sequence for $\phi=\pi$ cancels even for $2N=1$.

The sequence in \CHANGE{Fig.~\ref{Fig1} (b)} has a first order amplitude error that is only cancelled for $\phi=\pi$ (XY-sequence), and is therefore only robust against Rabi frequency errors for $\phi \cong \pi$.
The detuning errors affect the evolution as
\begin{equation}
U_\text{sequence, 2N} = U_\text{perfect-pulses} \times (1+\delta \begin{pmatrix}
i \cos(2N\phi) & \sin(N\phi) \\ \sin(N\phi) & -i \cos(2N\phi)
\end{pmatrix}).
\end{equation}
that is always bounded and does not increase with the number of pulses.

\subsection{Appendix B: Entangling by using polarization transfer Hamiltonians}
A simultaneous flipflop between an NV and 2 nuclei can be written as
\begin{figure*}[h!]
	\scalebox{.8}{\parbox{\linewidth}{%
			\begin{align}
			\nonumber&\exp \left(-i t/2 \left[ a_1(\sigma_x^{(NV)}\otimes I_x^{(1)} + \sigma_y^{(NV)}\otimes I_y^{(1)}) + a_2(\sigma_x^{(NV)}\otimes I_x^{(2)} + \sigma_y^{(NV)}\otimes I_y^{(2)})    \right]  \right)
			\\&=\left(
			\begin{array}{cccc | cccc}
			1 & 0 & 0 &0 &0 &0 &0 &0
			\\
			0 & \xa{\frac{a_1^2 + a_2^2 \cos(\sqrt{a_1^2+a_2^2}t/2)}{a_1^2+a_2^2}} & \xb{\frac{-2a_1a_2 \sin^2( \sqrt{a_1^2+a_2^2}t/4)}{a_1^2+a_2^2}} & 0 &
			\xc{\frac{-ia_2 \sin( \sqrt{a_1^2+a_2^2}t/2)}{\sqrt{a_1^2+a_2^2}}} &0 &0 &0
			\\
			0 & \xb{\frac{-2a_1a_2 \sin^2( \sqrt{a_1^2+a_2^2}t/4)}{a_1^2+a_2^2}} & \xa{\frac{a_2^2 + a_1^2 \cos(2 \sqrt{a_1^2+a_2^2}t/4)}{a_1^2+a_2^2}}  & 0 &
			\xc{\frac{-ia_1 \sin( \sqrt{a_1^2+a_2^2}t/2)}{\sqrt{a_1^2+a_2^2}}} &0 &0 &0
			\\
			0 & 0 & 0 &\xa{\cos(\sqrt{a_1^2+a_2^2}t/2))} &0 &\xc{\frac{-ia_1 \sin( \sqrt{a_1^2+a_2^2}t/2)}{\sqrt{a_1^2+a_2^2}}} &\xc{\frac{-ia_2 \sin( \sqrt{a_1^2+a_2^2}t/2)}{\sqrt{a_1^2+a_2^2}}} &0
			\\
			\hline
			\\
			0 &\xc{\frac{-ia_2 \sin( \sqrt{a_1^2+a_2^2}t/2)}{\sqrt{a_1^2+a_2^2}}} &\xc{\frac{-ia_1 \sin( \sqrt{a_1^2+a_2^2}t/2)}{\sqrt{a_1^2+a_2^2}}} &0 & \xa{\cos(\sqrt{a_1^2+a_2^2}t/2))} &0 & 0 & 0
			\\
			0 & 0 & 0  & \xc{\frac{-ia_1 \sin( \sqrt{a_1^2+a_2^2}t/2)}{\sqrt{a_1^2+a_2^2}}} &
			0 &\xa{\frac{a_2^2 + a_1^2 \cos(2 \sqrt{a_1^2+a_2^2}t/4)}{a_1^2+a_2^2}} &\xb{\frac{-2ab \sin^2( \sqrt{a^2+b^2}t)}{a^2+b^2}} &0
			\\
			0 & 0 & 0 & \xc{\frac{-ia_2 \sin( \sqrt{a_1^2+a_2^2}t/2)}{\sqrt{a_1^2+a_2^2}}} &
			0 &\xb{\frac{-2ab \sin^2( \sqrt{a^2+b^2}t)}{a^2+b^2}} & \xa{\frac{a_1^2 + a_2^2 \cos(\sqrt{a_1^2+a_2^2}t/2)}{a_1^2+a_2^2}} &0
			\\
			0 & 0 & 0 &0 &0 &0 &0 &1
			\end{array}
			\right)\nonumber
			\end{align}
	}}
\end{figure*}

where the solid-framed matrix elements are NV-nuclear interactions, dotted matrix elements are to preserve unitarity and dashed matrix elements are nuclear-nuclear interactions.
For $\sin( \sqrt{a_1^2+a_2^2}t/4) = 1$ and $a_1=a_2$, we obtain

\begin{align}
&\exp \left(-i t \left[ a_1(\sigma_x^{(NV)}\otimes I_x^{(1)} + \sigma_y^{(NV)}\otimes I_y^{(1)}) + a_2(\sigma_x^{(NV)}\otimes I_x^{(2)} + \sigma_y^{(NV)}\otimes I_y^{(2)})    \right]  \right)
\\&=\left(
\begin{array}{cccc | cccc}
1 & 0 & 0 &0 &0 &0 &0 &0
\\
0 & \xa{0} & \xb{-1} & 0 &
\xc{0} &0 &0 &0
\\
0 & \xb{-1} & \xa{0}  & 0 &
\xc{0} &0 &0 &0
\\
0 & 0 & 0 &\xa{-1} &0 &\xc{0} &\xc{0} &0
\\
\hline
\\
0 &\xc{0} &\xc{0} &0 & \xa{-1} &0 & 0 & 0
\\
0 & 0 & 0  & \xc{0} &
0 &\xa{0} &\xb{-1} &0
\\
0 & 0 & 0 & \xc{0} &
0 &\xb{-1} & \xa{0} &0
\\
0 & 0 & 0 &0 &0 &0 &0 &1
\end{array}
\right)\nonumber
\end{align}
This means if the NV is prepared in $\ket{1}$, the nuclear states will swap and a \mbox{CPHASE} gate is applied.

\subsection{Appendix C: Calculation of the effective coupling}

\begin{figure}[h!]
	\centering
	\includegraphics[width=.49\linewidth]{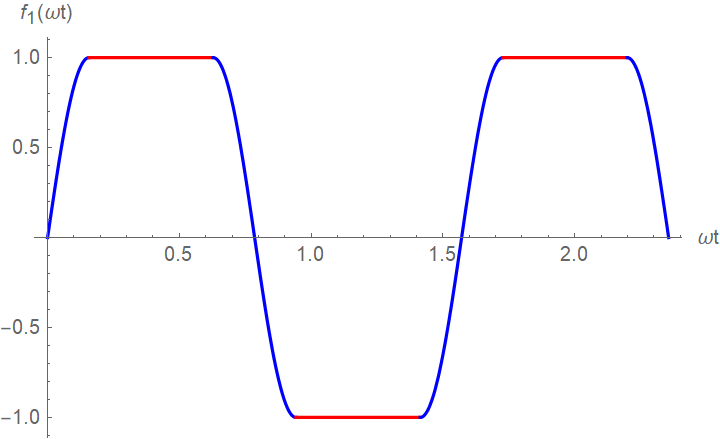}
	\caption{\CHANGE{The modulation} function $f_1$ for the adapted sequence until the center of the \CHANGE{central $\pi$-pulse ($t=\tau$)
		rises} to 1 during first $\pi$/2 pulse and remains there until the beginning of the composite pulse. For symmetry reasons it is sufficient to consider only the first $\pi$-pulse ($1\rightarrow-1$), the evolution afterwards that takes a fraction $\beta_1$, the second $\pi$-pulse ($-1\rightarrow1$), the evolution afterwards that takes a fraction $\beta_2$ and half of the third $\pi$-pulse ($1\rightarrow0$).}
	\label{FigSI-A}
\end{figure}

We analytically calculate the effective coupling constant for the sequence in \CHANGE{Fig.~\ref{Fig2} (a)} with finite pulses. 
\CHANGE{To do so, we describe the effective $S_z$ operator in \CHANGE{Eq.~}(\ref{eq10}) using the modulation function $f_1(t)$, following a similar path as for the special case of the PulsePol sequence \cite{PP}. 
	In particular, this modulation function equals 1 for every position that is marked with 1 in \CHANGE{Fig.~\ref{Fig2} (a)}, and -1 for every position 2, such that the effective operator is $f_1(t) \sigma_x/2 $. Positions 3 and the following are automatically included in the following discussion, with periodically changing phases.
During pulses, the modulation function is rotated similar to magnetisation, leading to

\begin{align}
	f_1(t)=
	\begin{cases}
		\sin(\Omega t), & \text{for } 0 \le t \le \frac{\pi}{2\Omega}\\
		1, &\text{for } \frac{\pi}{2\Omega} \le t < t_1-\frac{\pi}{2\Omega}\\
		-\sin(\Omega (t-t_1)), & \text{for } -\frac{\pi}{2\Omega} \le t-t_1 \le \frac{\pi}{2\Omega}\\
		-1, &\text{for } t_1+\frac{\pi}{2\Omega} \le t < t_2-\frac{\pi}{2\Omega}\\
		\sin(\Omega (t-t_2)), & \text{for } -\frac{\pi}{2\Omega} \le t-t_2 \le \frac{\pi}{2\Omega}\\
		1, &\text{for } t_2+\frac{\pi}{2\Omega} \le t < \tau-\frac{\pi}{2\Omega}\\
		-\sin(\Omega (t-\tau)), & \text{for } -\frac{\pi}{2\Omega} \le t-\tau \le 0\\
		-f_1(t-\tau), & \text{for } \tau \le t \le 2\tau\\
		f_1(t-2\tau), & \text{for } t \ge 2\tau\\
	\end{cases}
\end{align}
due to the $2\tau$-periodicity and antisymmetry with respect to the center of the third pulse $t=\tau$.
Here $t_1 = \tau - \tau_1-\tau_2-2\pi/\Omega$ and $t_2 = \tau -\tau_2-\pi/\Omega$ are the centers of the first and second $\pi$-pulse, respectively. This leads to the function in \CHANGE{Fig.} \ref{FigSI-A}.
}

\CHANGE{For symmetry reasons, it is sufficient to calculate the overlap of this function with $\sin(\omega(\tau-t))$, as the Fourier coefficient of $\cos(\omega(\tau-t))$ vanishes due to the point symmetry with respect to $t=\tau$. As the Fourier coefficient only gives the prefactor of the $S_xI_x$ term in a rotated basis, an additional factor $1/2$ is necessary to obtain the effective coupling for the flipflop Hamiltonian in \CHANGE{Eq.~(\ref{eq10a})}. This leads to

%It is sufficient to calculate the overlap of the modulation function with the intended cosine from the first $\pi/2$ pulse until the center of the first $\pi$-pulse as shown in Figure \ref{FigSI-A}.
%Mathematically this corresponds to calculating the Fourier coefficient for the desired resonance.
%We denote the fraction that the times $\tau_{1,2}$ between the additional pulses by $\beta_{1,2}$ such that $\tau_{1,2} = \beta_{1,2} \frac{\pi+\phi}{2\omega_L}$, see Figure \ref{FigSI-A}.
%For the resonance condition $(n, \pm)$ we denote $\phi_{n, \pm} = 2n\pi \pm \phi$ to obtain

%\begin{align}
%	a_\text{eff} &= \frac{a^\perp}{\pi+\phi_{n, \pm}}  \int\limits_{-\phi_{n, \pm}/2}^{\pi/2}  \mathrm{d}x \cos(x) f_1(x)
%	\\&= \frac{a^\perp}{\pi+\phi_{n, \pm}} \frac{1}{1-\epsilon^2}
%	\left[ \epsilon \cos \frac{\phi_{n, \pm}}{2}  + \cos \frac{\epsilon\pi}{2}  + \cos \frac{(\beta_1+\beta_2)(\phi_{n, \pm} + \pi) + 3\epsilon\pi}{2}
%	+ \cos \frac{(\beta_1+\beta_2)(\phi_{n, \pm} + \pi) + 5\epsilon\pi}{2}
%	\right.\\&\hspace{1 cm}\left.
%	- \cos \frac{\beta_2(\phi_{n, \pm} + \pi) + \epsilon\pi}{2}   - \cos \frac{\beta_2(\phi_{n, \pm} + \pi) + 3\epsilon\pi}{2}     + \sin \frac{\phi_{n, \pm} - \epsilon\pi}{2}                    \right]
%	%
%\end{align}

\begin{align}
\nonumber	a_\text{eff} &= \frac{a^\perp}{2\tau}  \int\limits_{0}^{\tau}  \mathrm{d}t \sin(\omega(\tau-t)) f_1(t)
	\\&= \frac{a^\perp}{2\tau\omega_L} \frac{1}{1-\epsilon^2}
	\left[ - \cos \left(\frac{\epsilon\pi}{2} -\omega_L \tau \right) + \cos \left(\frac{\epsilon\pi}{2}\right)
		\left(1-\epsilon^2+\epsilon^2 \cos \left(\frac{(\tau_1+\tau_2)\omega_L}{\epsilon}\right)
		-2 \cos \left(\epsilon\pi + \tau_2 \omega_L\right) \nonumber
		\right.\right. \\& \nonumber \left.\left. \hspace{1cm}
		+2 \cos \left(2\epsilon\pi + (\tau_1+\tau_2) \omega_L\right)
		\right)
		+ \epsilon \left(\sin(\omega_L \tau) + \left(\epsilon-\sin\left(\frac{\epsilon\pi}{2}\right) \sin\frac{(\tau_1+\tau_2)\omega_L}{\epsilon}\right)\right)
		\right]
		\\&= \nonumber
		\frac{a^\perp}{2\tau\omega_L} \left[1-\cos(\omega_L \tau) -2\cos(\omega_L \tau_2) +2\cos(\omega_L (\tau_1+\tau_2)) \right] + O(\epsilon)
		\\&= \frac{a^\perp}{2\tau\omega_L} \left[1-\cos(\omega_L \tau)  \right] + O(\epsilon) + O(\omega_L \tau_1)  + O(\omega_L \tau_2)	
%		\\&
%	+ \cos \frac{\epsilon\pi}{2}  + \cos \frac{(\beta_1+\beta_2)(\phi_{n, \pm} + \pi) + 3\epsilon\pi}{2}
%	+ \cos \frac{(\beta_1+\beta_2)(\phi_{n, \pm} + \pi) + 5\epsilon\pi}{2}
%	\right.\\&\hspace{1 cm}\left.
%	- \cos \frac{\beta_2(\phi_{n, \pm} + \pi) + \epsilon\pi}{2}   - \cos \frac{\beta_2(\phi_{n, \pm} + \pi) + 3\epsilon\pi}{2}     + \sin \frac{\phi_{n, \pm} - \epsilon\pi}{2}                    \right]
	%
\end{align}
}

where we defined $\epsilon = \omega_L/\Omega$ as the ratio between nuclear Larmor and electron Rabi frequency. \CHANGE{Note that the resonance $(n, \pm)$ and the phase $\phi$ are included in $\omega_L \tau$ according to \CHANGE{Eq.~(\ref{gateeqAX})}.
	We obtain
 the expected $a^\perp(2+\sqrt{2})/(3\pi)$ for PulsePol (n=0,+; $\tau = 3\pi/4\omega_L$) with instantaneous pulses ($\epsilon=\beta_1=\beta_2=0$).}

\begin{figure}[h!]
	\centering
	\includegraphics[width=.49\linewidth]{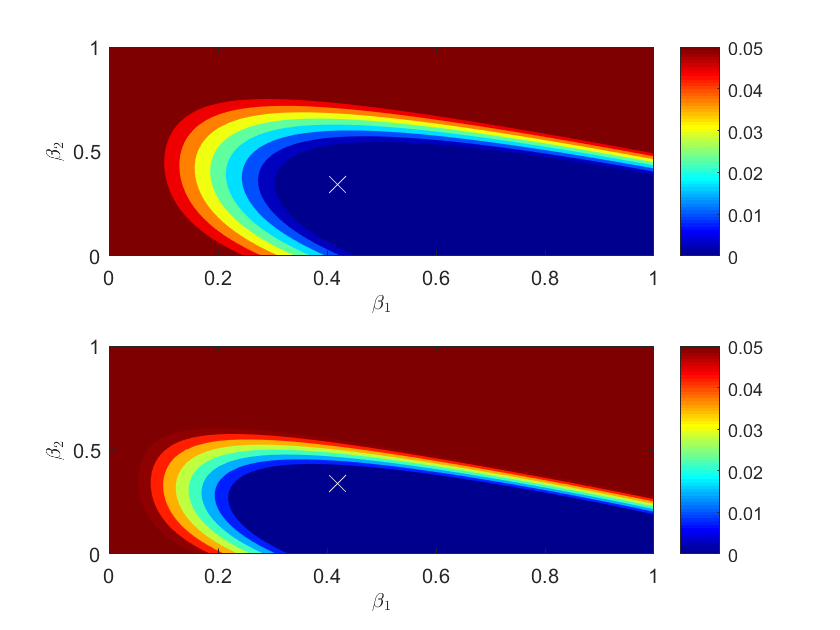}
	\includegraphics[width=.49\linewidth]{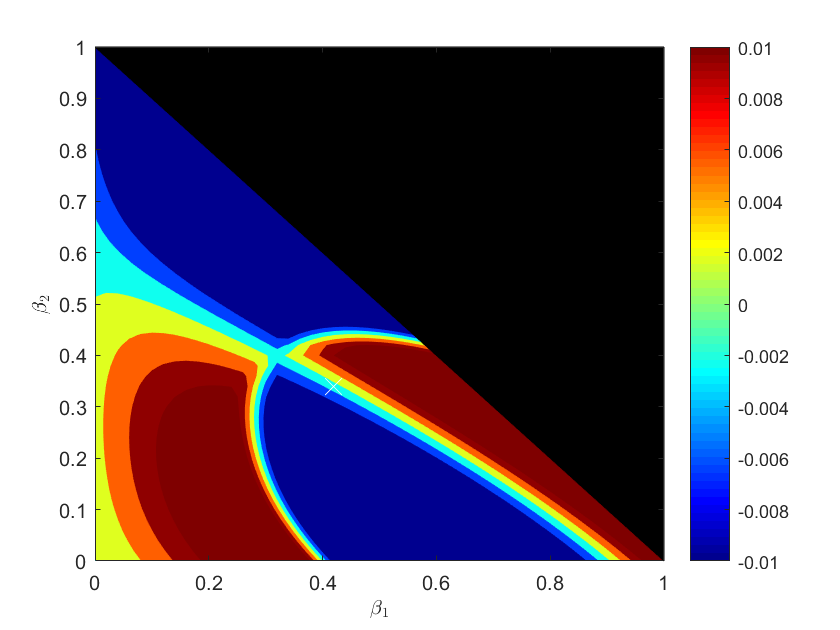}
	\caption{Comparison of effective coupling (left) for $A_\text{eff}(\phi_1, \beta_1, \beta_2)$ (top) and $ A_\text{eff}(2\pi-\phi_1, \beta_1, \beta_2)$ (bottom) for the parameters in \CHANGE{Fig.} \ref{Fig2} (main text) and difference between those functions (right). \CHANGE{Here $\tau_{1,2} = \beta_{1,2} \frac{\pi+\phi}{2\omega_L}$ was used as dimensionless time $\beta_i<1$}. The chosen $\beta_{1,2}$ is marked with a white cross where bot functions were equal and close to a desired value.}
	\label{FigSI-B}
\end{figure}

This equation can be used to achieve the same coupling for two different resonances. 
We use this equation to calculate equal coupling for the $n=0,+$ and $n=1,-$ resonances as shown in \CHANGE{Fig.} \ref{FigSI-B}.

\end{document}